\documentclass{article}
\usepackage{graphicx,amsmath,amssymb,amsfonts,amsthm}
\begin{document}

\title{On black holes as inner boundaries for the constraint equations}

\author{\it S. Dain \\
\rm Max-Planck-Institut f\"ur Gravitationsphysik\\
  Am M\"uhlenberg 1\\
  14476 Golm\\
  Germany\\
e-mail: dain@aei.mpg.de}

\maketitle

\abstract{General aspects of the boundary value problem for the
  constraint equations and their application to black holes are
  discussed.}

\section{Introduction}  \label{intro}
There are many kinds of boundaries in physics. For example, in
electrodynamics, the boundary given by the interface between a charge
distribution and vacuum. However, this is not a fundamental boundary
of Maxwell equations, in the following sense.  This boundary is
introduced in the sources by choosing a charge density with compact
support.  The sources satisfy extra equations.  Maxwell equations are
expected to be fundamental equations; matter sources equations are
phenomenological approximation suitable to describe some specific
matter models.  The same apply to matter sources in Einstein
equations. In this case the space time boundary is introduced in the
sources of Einstein equations by choosing an energy density with
compact support.

In the case of vacuum Maxwell equations, the field does not interact
with it self, hence it can not produce ``its own boundary''.  In the
case of Einstein vacuum equation there exists such fundamental kind of
boundary produced only by the self interaction of the vacuum field:
black holes.

In the context of an initial value formulation, the first step in
order to understand the space time boundary produced by a black hole
is the study of the intersection of this boundary with a spacelike
three-dimensional Cauchy hypersurface. That is, to study the black
hole boundary value problem for the constraint equations.  This
problem has been recently studied in \cite{Maxwell03} \cite{Dain03}.

Since only fundamental properties of gravity are involved in a vacuum
black hole, it can be expected that black hole boundary conditions can
be written in a geometrical form. It turns out that this is true.
Moreover, black holes boundaries for the constraint equations suggest
a deep interplay between Riemannian geometry, elliptic equations and
physics. In the present article we discuss some general aspects about
this interplay.

In section \ref{cons} we present the constraint equations and the
corresponding boundary conditions.  In section \ref{elliptic} we
discuss elliptic reductions to this equation.  General kind of
boundary conditions are discussed in section \ref{boundary}. In
section \ref{phys} we discuss boundary conditions that are physically
meaningful. Finally, in section \ref{bh}, we discuss black hole
boundary conditions.

\section{The constraint equations}
\label{cons}

Let $C_k$ be a finite collection of \emph{compact} sets in
$\mathbb{R}^3$. We define the \emph{exterior region} $\tilde
\Omega=\mathbb{R}^3\setminus \cup_k C_k $. For simplicity, we will
mainly consider the constraint equations in the time-symmetric case,
that is, when extrinsic curvature vanishes. However, all the following
consideration apply also to the general case (see \cite{Dain03}).  Let
$\tilde h_{ab}$ be a Riemannian metric on $\tilde \Omega$ and let
$\tilde R$ be the corresponding Ricci scalar.  The time-symmetric,
vacuum, constraint equation is given by
\begin{equation}
  \label{eq:8}
  \tilde R = 0,
\end{equation}
on $\tilde \Omega$.

There exists two kinds of boundary conditions for equation
\eqref{eq:8} in $\tilde \Omega$: outer boundary conditions and inner
boundary conditions. The outer boundary condition is asymptotic
flatness, and it is essentially a fall of condition on $\tilde
h_{ab}$.  Physically, it means that we have an isolated system.  The
data will be called \emph{asymptotically flat} if there exists some
compact set $C$, with $\cup_k C_k \subset C$, such that $\tilde \Omega
\setminus C$ can be mapped by a coordinate system $\tilde x^j$
diffeomorphically onto the complement of a closed ball in
$\mathbb{R}^3$ and we have in these coordinates
\begin{equation} 
\label{pf1}
\tilde h_{ij}=(1+\frac{2m}{\tilde r})\delta_{ij}+O(\tilde r^{-2}),
\end{equation}
as $\tilde r= ( \sum_{j=1}^3 ({\tilde x^j})^2 ) ^{1/2} \to \infty$,
where the constant $m$ is the total mass of the initial data.

The inner boundary condition will be the black hole boundary
condition.  The boundaries $\partial C_k$ are assumed to be smooth, two
dimensional surfaces in $(\tilde \Omega, \tilde h )$. Let $\tilde
\nu^a$ be the unit normal of $\partial C_k$, with respect to $\tilde
h_{ab}$, pointing in the \emph{outward} direction of $\tilde \Omega$.
Let $t^a$ be the unit timelike vector field orthogonal to the
hypersurface $\tilde \Omega$ with respect to the spacetime metric
$g_{ab}$ ($t^at^b g_{ab}=-1$ with our signature convention) The
outgoing and ingoing null geodesics orthogonal to $\partial C_k$ are
given by $ l^a=t^a-\tilde \nu^a $ and $k^a=t^a+\tilde \nu^a$
respectively, the corresponding expansions are given by
$\Theta_+=\nabla_a l^a$ and $\Theta_-=\nabla_a k^a$, where $\nabla_a$
is the connexion with respect to $g_{ab}$.  We
can calculate these expansions in terms of quantities intrinsic to the
initial data. In the particular case of time-symmetric data we have
\begin{equation}
  \label{eq:21}
  \Theta_- =-\Theta_+= \tilde H, 
\end{equation}
where $\tilde H = \tilde D_a\tilde \nu^a$, and $\tilde D_a$ is the
covariant derivative with respect $\tilde h_{ab}$. That is, $\tilde
H$ is the mean curvature of the two-dimensional surface $\partial
\tilde \Omega $ with respect to the metric $\tilde h_{ab}$ and the
normal $\tilde \nu^a$.  The boundary will be called \emph{marginally
  trapped} if
\begin{equation}
\label{eq:min}
\tilde H =0.
\end{equation}
A marginally trapped surface indicates the presence of a black hole
(see \cite{Wald84} and also the discussion in \cite{Dain03}).
Equation \eqref{eq:min} geometrically means that the two dimensional
boundary is an extremal surface with respect to the metric $\tilde
h_{ab}$.  Equation \eqref{eq:min} will be our inner boundary
condition.

\section{The constraint equations as an elliptic system}
\label{elliptic}
 
We want to find solutions $\tilde h_{ab}$ to equation \eqref{eq:8}
which satisfy boundary conditions \eqref{pf1} and \eqref{eq:min}.  If
we write equation \eqref{eq:8} in terms of the metric components of
$\tilde h_{ab}$ we get a complicated non linear equation. We have six
unknown functions in the metric $\tilde h_{ab}$ and only one equation.
That is, we have an underdetermined system.  The strategy to solve
this equation is to split the six unknowns into two sets.  One set
will be called the \emph{free data}, we want to prescribe them freely
or at least with some restrictions easy to achieve, which, in
particular, do not involve solving differential equations. Roughly
speaking, the free data set should contain five free functions. The
other set will contain only one function. For a given choice of free
data we have to solve equation \eqref{eq:8} to calculate this
function.
 
There exists a priori many ways of doing this splitting. For example
we can chose some coordinate system and chose the free data set to be
five components of the metric in these coordinates and try to solve
the equation for the remainder.  But the resulting equation will be in
general a complicated non linear equation which is of no known type
and hence will be difficult, if not impossible, to prove that in fact
for every choice of free data we do get a solution for the remainder.
 
In order to control the behavior of the solution $\tilde h_{ab}$ at
the boundary we need also to prescribe boundary conditions for the
unknown function.  Our splitting and boundary conditions will be
successful if for arbitrary free data and boundary conditions we
always get a solution for the reminder. This suggests that the problem
has an elliptic nature. To some extend this is true, the constraint
equation can be reduced to an elliptic system. However, this is not
the only way of solving them, for example they can be reduced to a
parabolic system \cite{bartnik93} and this lead to the discovery of
new kinds of solutions.  Nevertheless, so far only the elliptic approach has
been successful in the general case (the parabolic system has been
mainly studied in the time symmetric case).

There is not a unique way of getting an elliptic system out of
equation \eqref{eq:8}. Different elliptic systems will lead to
different choices of free data and boundary conditions, and hence they
can be more appropriate to describe different kinds of physical
situations.  Here we will use the so called conformal method, which is
probably the simplest one, since it lead to semilinear equations,
which reduce to a linear equation in the time-symmetric case. Other
elliptic reductions give quasi linear equations \cite{Corvino99}
\cite{Corvino:2003sp} \cite{Chrusciel:2003sr}
\cite{Chrusciel:2002vb}\cite{Butscher:2002pq}.  So far, black hole
boundary conditions have been only studied using the conformal method
\cite{Maxwell03} \cite{Dain03}.  It will very be interesting to study
them with other elliptic reductions.

The conformal method is as follows. Let $h_{ab}$ be a Riemannian
metric, let $\psi$ be a \emph{positive} solution of the following
equation
\begin{equation}
\label{Lich}
L_h \psi=0, 
\end{equation}
where $L_h\equiv D^aD_a-R/8$, $D_a$ is the covariant derivative with
respect to $h_{ab}$ and $R$ is the Ricci scalar of the metric
$h_{ab}$.  Then, the rescaled metric $\tilde h_{ab}=\psi^4 h_{ab}$
will satisfy equation \eqref{eq:8}. Note that the differential
operator $L_h$ is elliptic, hence the linear equation \eqref{eq:8} is
an elliptic reduction to the time symmetric constraint equation
\eqref{eq:8}.

Two metrics $\hat h_{ab}$ and $h_{ab}$ belong to the same conformal
class if there exists a positive conformal factor $\hat \psi$ such
that $\hat h_{ab}=\hat \psi^4 h_{ab}$. For any metric in the same
conformal class we get the same solution $\tilde h_{ab}$. That is, the
free data set is given by the conformal class of metrics, which can be
represented by five free functions.

\section{Elliptic boundary conditions}
\label{boundary}

What kind of boundary conditions are compatible with the constraint
equations? If we have reduced them to an elliptic system, this
question can be answer in full generality.  Given an elliptic system,
the boundary conditions will lead to a well posed problem if and only
if they satisfy the so called \emph{complementing condition} or
\emph{Lopatinski-Schapiro} conditions (see \cite{Agmon59}
\cite{Agmon64} for a precise statement and also the introductory
book \cite{treves:75}). These are conditions at the linear level.  For
non linear system, these conditions are imposed to the associated
linearized problem. Of course in the non linear case, these conditions
are in general only necessary but not sufficient to prove the existence
of solution. For non linear systems we have to study each particular
case to decide whether there exists solutions or not.

Simple examples of boundary conditions that satisfy the
Lopatinski-Schapiro requirements for equation \eqref{eq:8} are Dirichlet
and Neumann boundary conditions for the conformal factor $\psi$. More
general boundary conditions are possible, it is even possible that the
order of the derivatives in the boundary operator is higher than the
order of the derivatives in the differential operator.

However, here we are interested in positive solutions. This is an
extra requirement of our particular elliptic reduction. If $\psi$ is
zero at some point then $\tilde h_{ab}= \psi^4h_{ab}$ will not be a
Riemannian metric at this point.  The positivity of the solution can
be proved by a particular feature of second order elliptic equations:
the maximum principle (see, for example, \cite{Gilbarg}).
 
In order to use the maximum principle we need an extra requirement on the
lower order coefficients of the operator $L_h$
\begin{equation}
  \label{eq:10}
  R\geq 0.
\end{equation}

The most general kind of boundary condition that satisfy the
Lopatinski-Shapiro conditions and also allow us to use the maximum
principle to prove positivity is given by the Dirichlet boundary condition
\begin{equation}
  \label{eq:9}
  \psi=\varphi \quad \text{ on } \partial \tilde \Omega
\end{equation}
or the oblique derivative boundary condition
\begin{equation}
\label{eq:9b}
\beta^a D_a \psi + \alpha \psi =\varphi \quad \text{ on } \partial
\tilde \Omega,
\end{equation}
where $\varphi$, $\alpha$ and $\beta^a$ are arbitrary
functions which satisfy $\varphi\geq0$, $\alpha \geq 0$ and $\beta^a
\nu_a>0$ on $\partial \tilde \Omega$, where $\nu^a$ is the outward
unit normal with respect to the conformal metric $h_{ab}$. We note
that $\beta^a\nu_a>0$ guarantee that \eqref{eq:9b} satisfies the
Lopatinski-Schapiro conditions.

In the exterior region $\tilde \Omega$ we need in addition the
asymptotic flatness condition which in this case is given by
\begin{equation}
  \label{eq:2}
  \lim_{r \to \infty} \psi =1.
\end{equation}

Conditions \eqref{eq:10}, \eqref{eq:9} or \eqref{eq:9b}, \eqref{eq:2}
will guarantee the existence of a unique positive solution $\psi$ of
equation \eqref{Lich}, and hence an asymptotically flat metric $\tilde
h_{ab}$ which satisfies the constraint equation \eqref{eq:8} with
\eqref{eq:9} or \eqref{eq:9b} as inner boundary condition.  Of course,
for arbitrary $\varphi$, $\alpha$ and $\beta^a$, the boundary condition
\eqref{eq:9b} will not have any interesting geometrical
meaning.

\section{On physical boundary conditions}
\label{phys}
Probably any smooth solution of the constraint equation in some bounded region
can have some physical interpretation in the sense that this bounded
region can be a piece of a space time that can describe some physical
phenomena.  The situation is different in the case of an exterior
region $\tilde \Omega$: in order to be physically meaningful as a
description of an isolated system the solution must have positive
mass. Since we  artificially cut out a region in $\mathbb R^3$,
the positivity mass theorem does not automatically apply. Many of the
solutions found in the previous section will have negative total mass.
This can be explicitly seen in the following example.

Let us consider the Schwarzschild, time symmetric, initial data.  In
this case $h_{ab}=\delta_{ab}$, where $\delta_{ab}$ is the flat
metric, $L_h=\Delta$, where $\Delta$ is the flat Laplacian and the
conformal factor is given by
\begin{equation}
  \label{eq:4}
  \psi = 1+\frac{m}{2r}.
\end{equation}
We chose the exterior region $\tilde \Omega$ to be  the exterior of a ball
of radius $r=a$. Note that with our conventions
$\nu^a=-(\partial/\partial r)^a$.  The Dirichlet boundary condition is
given by
\begin{equation}
  \label{eq:5}
  \psi = \varphi_0  \quad \text{ on } \partial \tilde \Omega,
\end{equation}
where $\varphi_0$ is a positive constant.  Using \eqref{eq:4} one
easily check that $\varphi_0<1$ implies $m<0$ and $\varphi_0>1$
implies $m>0$. Take $a>m/2$, then we will have $\psi>0$ in the
exterior region and $m<0$ if we chose $\varphi_0<1$. This means that
$\varphi_0<1$ is not a physical boundary condition although it
mathematically consistent in the sense that it gives us existence and
uniqueness of a positive solution.  In the general case, if we impose
Dirichlet boundary conditions to the conformal factor appears to be
difficult to recognize for which value of the boundary function we
will get data with positive mass. We conclude that Dirichlet boundary
conditions for $\psi$ are not physically meaningful in general.

In this example, the Neumann condition
\begin{equation}
  \label{eq:7}
  \nu^aD_a \psi = - \left. \frac{\partial \psi}{\partial r}\right |_{r=a}=\varphi_0 
\end{equation}
with $\varphi_0>0$ always produce data with positive mass, since
$m=2\varphi_0a^2$.  This can be generalized. Consider the case of non
trivial extrinsic curvature, and assume that the trace of it is zero.
Assume that the conformal metric $h_{ab}$ satisfies $R=0$ and the
following fall off $h_{ij}=\delta_{ij}+O(r^{-2})$. The equation
for the conformal factor is given by
\begin{equation}
  \label{eq:11}
  L_h\psi = D^aD_a \psi = -\frac{K_{ab}K^{ab}}{8\psi^7},
\end{equation}
where $K_{ab}$ is the rescaled extrinsic curvature. 
Let us impose Neumann boundary condition to this equation
\begin{equation}
  \label{eq:1}
\nu^aD_a \psi =\varphi \quad \text{ on } \partial \tilde \Omega, 
\end{equation}
with $\varphi \geq 0$. Integrating equation \eqref{eq:11} on $\tilde
\Omega$ and using the Gauss theorem  we get
\begin{equation}
  \label{eq:12}
  m\geq \int_{\partial \tilde \Omega} \nu^a D_a \psi \, dS\geq 0,
\end{equation}
where we have used \eqref{eq:1} and  the following expression for the mass
\begin{equation}
  \label{eq:13}
  m=- \frac{1}{4\pi}\lim_{r\to \infty} \int_{S_r} \frac{\partial \psi}{\partial
    r} \, dS_r.
\end{equation}

We conclude that every data which satisfy the boundary condition
\eqref{eq:1} in the appropriate conformal class will have positive
mass.  Note that the boundary condition \eqref{eq:1} is linear even in
the general, non time-symmetric, case. It is not clear to me the
meaning of this condition, and I am not aware of any application of it.
Although in \cite{Maxwell03} only black hole boundary conditions has
been studied, remarkably, all the solutions founded there satisfy in
addition condition \eqref{eq:1}.

\section{Black hole boundary conditions}
\label{bh}
The most important inner boundary condition for the vacuum constraint
equation is the black hole condition \eqref{eq:min}. This boundary
condition has both a geometric and physical meaning. There exist
versions of the positivity mass theorem which include black holes as
inner boundaries (see for example \cite{reula:84}).

Condition \eqref{eq:min} can
be written in terms of the conformal quantities as
\begin{equation}
\label{eq:con-min}
4\nu^aD_a\psi  + H\psi=0, 
\end{equation}
where $H=D_a\nu^a$.  

If $H\geq 0$ on $\partial \tilde \Omega$, then condition
\eqref{eq:con-min} has the form \eqref{eq:9b}.  If, in addition, the
conformal metric $h_{ab}$ satisfies $R\geq 0$ on $\tilde \Omega $ we
can apply the standard linear elliptic theory and the maximum
principle to prove that there will exist a unique positive solution
$\phi$ of our problem.

The conformal metric is not really a free data, it should
satisfy the condition $R\geq 0$, and also the boundary $\partial \tilde
\Omega$ should satisfy $H\geq 0$. Note that if we chose the conformal
metric to be the flat metric and  $\partial \tilde
\Omega$ any sphere, then this boundary will have $H < 0$, and hence it
does not satisfy our hypothesis. However, it is simple to construct
families of metric which satisfy  $R\geq 0$ and $H\geq 0$ on $\partial \tilde
\Omega$. Consider, for example, the time-symmetric initial data for
Reissner-Nordstrom. The metric is given by
\begin{equation}
  \label{eq:3}
  h_{ab}=\hat \psi^4 \delta_{ab},
\end{equation}
where 
\begin{equation}
  \label{eq:6}
  \hat \psi= \frac{1}{2r} \sqrt{(q+2r+m)(-q+2r+m)}.
\end{equation}
The constants $q$ and $m$ are the charge and the mass of the data
respectively. We  assume $q^2<m^2$. The Ricci scalar is given by
\begin{equation}
  \label{eq:14}
  R=\frac{2q^2}{\hat\psi^8 r^4}, 
\end{equation}
which is positive.  For $r < r_0=(\sqrt{m^2-q^2})/2$ the mean
curvature of the two surfaces of constant radius is positive. Then
this metric satisfies our hypothesis. Moreover, take $q>0$ (this
implies $R>0$), let $\partial \tilde \Omega$ such that $H>0$, let
$\epsilon$ small enough and $\lambda_{ab}$ and arbitrary tensor field;
then the metric $h_{ab}+\epsilon \lambda_{ab}$ satisfies also $R>0$
and $\partial \tilde \Omega$ will satisfy $H>0$ with respect to this
metric.

In the general case, black hole boundary conditions are non linear. 
This introduces extra difficulties in both the
existence proof and the election of free data. This problem has been
recently studied in \cite{Maxwell03} \cite{Dain03}.  In those
references, large classes of black hole exterior regions have been
constructed. However, it is still an open problem how to construct and
characterize \emph{all} possible initial data for black holes exterior
regions.


\begin{thebibliography}{10}

\bibitem{Agmon59}
S.~Agmon, A.~Douglis, and L.~Niremberg.
\newblock Estimates near the boundary for solutions of elliptic partial
  differential equations satisfying general boundary conditions {I}.
\newblock {\em Comm. Pure App. Math.}, 12:623--727, 1959.

\bibitem{Agmon64}
S.~Agmon, A.~Douglis, and L.~Niremberg.
\newblock Estimates near the boundary for solutions of elliptic partial
  differential equations satisfying general boundary conditions. {II}.
\newblock {\em Comm. Pure App. Math.}, 17:35--92, 1964.

\bibitem{bartnik93}
R.~Bartnik.
\newblock Quasi-spherical metrics and prescribed scalar curvature.
\newblock {\em J. Differential Geom.}, 37(1):31--71, 1993.

\bibitem{Butscher:2002pq}
A.~Butscher.
\newblock Perturbative solutions of the extended constraint equations in
  general relativity.
\newblock 2002, gr-qc/0211037.

\bibitem{Chrusciel:2002vb}
P.~T. Chrusciel and E.~Delay.
\newblock Existence of non-trivial, vacuum, asymptotically simple space-times.
\newblock {\em Class. Quant. Grav.}, 19:L71, 2002, gr-qc/0203053.

\bibitem{Chrusciel:2003sr}
P.~T. Chrusciel and E.~Delay.
\newblock On mapping properties of the general relativistic constraints
  operator in weighted function spaces, with applications.
\newblock 2003, gr-qc/0301073.

\bibitem{Corvino99}
J.~Corvino.
\newblock Scalar curvature deformation and a gluing construction for the
  einstein constraint equations.
\newblock {\em Commun. Math. Phys.}, 214(1), 1999.

\bibitem{Corvino:2003sp}
J.~Corvino and R.~M. Schoen.
\newblock On the asymptotics for the vacuum einstein constraint equations.
\newblock 2003, gr-qc/0301071.

\bibitem{Dain03}
S.~Dain.
\newblock Trapped surfaces as boundaries for the constraint equations.
\newblock 2003, gr-qc/0308009.
\newblock To appear in Class. Quantum Grav.

\bibitem{Gilbarg}
D.~Gilbarg and N.~S. Trudinger.
\newblock {\em Elliptic Partial Differential Equations of Second Order}.
\newblock Springer-Verlag, Berlin, 1983.

\bibitem{Maxwell03}
D.~Maxwell.
\newblock Solutions of the {E}instein constraint equations with apparent
  horizon boundary.
\newblock 2003, gr-qc/0307117.

\bibitem{reula:84}
O.~Reula and K.~P. Tod.
\newblock Positivity of the {B}ondi energy.
\newblock {\em J. Math. Phys.}, 25(4):1004--1008, 1984.

\bibitem{treves:75}
F.~Treves.
\newblock {\em Basic Linear Differential Equations}.
\newblock Pure and Applied Mathematics. Academic Press, New York, 1975.

\bibitem{Wald84}
R.~M. Wald.
\newblock {\em General Relativity}.
\newblock The University of Chicago Press, Chicago, 1984.

\end{thebibliography}

\end{document}